# Remarks on singular hypersurfaces and thin shells in general relativity

## Phillip W. Dennis[1]

Intelligent Systems Research, Thousand Oaks, CA

**Abstract** In this note several formulae that follow from W. Israel's [1] thin shell equation of motion (EoM) are derived. We review the main results of Israel's seminal papers [1,2]. The Israel EoM is solved for the three regimes of the total energy, *E<0, E=0* and *E>0*. This family of solutions is analogous to the three classes of solutions of the Lemaitre-Tolman-Bondi (LTB) equations [3]. We also derive the constraint that the gravitational mass is always positive. The main result of this note is the expressions for the interior and exterior metrics in terms of proper time on the shell. In these coordinates the time dilation formula is readily derived. We also derive the time dilation formula by direct geometric analysis. In particular we show that there is a differential time dilation of clocks at rest inside the cavity and clocks at rest just outside the cavity.

## I. Introduction

In this brief note we will examine features of the Israel thin shell solutions. The Israel EoM is solved for the three energy domains, *E<0, E=0* and *E>0*. We express the solution for the spacetime in terms of the shell proper time, and thereby derive the general formula for the time dilation between inertial clocks in the interior of the shell and outside the shell.

The paper employs natural units for which *G=c=1*. Also, $d\Omega^2 = d\theta^2 + \sin^2\theta d\phi^2$ denotes the metric on the unit sphere.

## II. Metric of the thin shell solution

As shown in [1,2], the space time can be covered by two coordinate charts.
$$\begin{aligned} ds^2 &= -dT^2 + dr^2 + r^2 d\Omega^2 \qquad &\text{if } r < R(\tau) \\ ds^2 &= -f dt^2 + f^{-1} dr^2 + r^2 d\Omega^2 \qquad &\text{if } r > R(\tau) \end{aligned} \qquad (1)$$

We have used $f(r) = (1 - 2M/r)$. The interior is flat Minkowski space and the exterior is the Schwarzschild solution. Both regions follow directly from Birkhoff's theorem.

The gravitating shell is the boundary at *R* between the interior cavity and the exterior.
$$ds^2 = -d\tau^2 + R(\tau)^2 d\Omega^2$$
Israel [1] derived the equation of motion of the free-falling shell. A first integral of that equation is [2]:
$$M = \mu(1 + \dot{R}^2)^{1/2} - \frac{\mu^2}{2R}. \qquad (2)$$

---

[1] Current email: pwdennis@earthlink.net

$$\dot{R} = \frac{dR}{d\tau}$$

$$M > 0$$

In this equation $\mu$ is the invariant mass that appears in the stress tensor; $M$ is the gravitational mass. The ratio $a = M/\mu$ appears as a constant of integration[2]. The mass $M$ includes the gravitational binding energy. It is to be noted that the range of the gravitational mass $M$ is strictly positive as equation (2) does not follow if $M=0$. This can be recognized by following the derivation of the EoM in Israel [1]. For the case $M= 0$ the interior and exterior of the shell is flat Minkowski space and the EoM in Israel reduces to:

$$2\ddot{R}\left(1+\dot{R}^2\right)^{1/2} = 0.$$

This yields the EoM:

$$\ddot{R} = 0,$$

implying $\dot{R} = constant$ or $R(\tau) = R_0 + \dot{R}_0 \tau$. This solution is consistent with an empty universe with no gravitational mass. It is, however, inconsistent with the limiting case $M \to 0$ of equation (2) which requires a static shell with $\dot{R}(\tau) = 0$ and $R_0 = \mu/2$. This suggest that $M=0$ requires $\mu = 0$ also. That this is the case can be seen from integrating the Einstein field equations. Using equation (83), p. 273, in Synge [4] leads to:

$$e^{-\alpha} = 1 - \frac{2M}{r} = 1 + \frac{8\pi}{r}\int_0^r r^2 T_4^4 dr.$$

The $T_4^4$ component of the stress-energy for a shell at rest at $R_0 = \mu/2$ is

$$T_4^4 = -\frac{\mu \delta(r-R_0)}{4\pi r^2}.$$

Substitution then yields

$$M = \mu,$$

thereby verifying that $M=0$ requires $\mu = 0$, which is the empty universe solution. Thus, $M > 0$.

A useful way of viewing equation (2) is as the expression for the total gravitational mass as a function of velocity and binding energy.

Equation (2) can be solved to give $\dot{R}$ in terms of $R$:

$$\dot{R}^2 = \left(\frac{M}{\mu} + \frac{\mu}{2R}\right)^2 - 1. \qquad (3)$$

From equation (3) we can derive the following constraint.

$$\dot{R}^2 = \left(\frac{M}{\mu} + \frac{\mu}{2R}\right)^2 - 1 \geq 0$$

Implying:

$$\left(\frac{M}{\mu} - 1\right) R \geq -\frac{\mu}{2}.$$

---

[2] Equation (2) corrects the typographical error in equation (21.176e) MTW [2], p. 556.

If $M \geq \mu$ (corresponding to the free or unbound shell) there is no constraint on the range of $R$. While, if $M < \mu$ (corresponding to a bound shell) we obtain the maximum radius of the shell:

$$R \leq \frac{\mu^2}{2(\mu - M)} \equiv R_{max}.$$

## III. Review of Solutions of the Israel EoM

In this section we review the solutions of the Israel EoM (equation (2) above) for the thin shell. Although Israel has solved them in parametric form in [5], we solve them herein as an implicit function of $R(\tau)$. Using equation (3) above we find:

$$\dot{R}^2 = \left(\frac{M^2}{\mu^2} - 1\right) + \frac{M}{R} + \frac{\mu^2}{4R^2}$$

$$= \frac{\left(\frac{M^2}{\mu^2} - 1\right)R^2 + MR + \frac{\mu^2}{4}}{R^2}$$

Giving

$$\frac{dR}{d\tau} = \frac{1}{R}\left[\left(\frac{M^2}{\mu^2} - 1\right)R^2 + MR + \frac{\mu^2}{4}\right]^{1/2}.$$

From this we obtain the following integral to be solved for $R(\tau)$:

$$\frac{RdR}{\left[\left(\frac{M^2}{\mu^2} - 1\right)R^2 + MR + \frac{\mu^2}{4}\right]^{1/2}} = d\tau. \qquad (4)$$

The coefficient of $R^2$ in equation (4) is a parameter that characterizes the energy state of the shell:

$$E = \left(\frac{M^2}{\mu^2} - 1\right).$$

There are three cases to consider.

If $-1 < E < 0$ we have a bound state which occurs when $0 < M < \mu$. This case occurs when the gravitational binding energy is positive resulting in a lower gravitational mass.

If $E = 0$ we have a free state which occurs when $M = \mu$. This case is when the gravitational binding energy is zero.

If $E > 0$ we have a free state which occurs when $M > \mu$.

These three cases correspond to the usual regimes as found in the LTB equations of motion [3].

Evaluating the integrals, we obtain implicit equations for $R$ (with initial condition $R_0 = R(\tau_0)$):

For positive *binding* energy, $E<0$:
$$\frac{\tau - \tau_0}{\mu} = F\left(\frac{R}{\mu}\right) - F\left(\frac{R_0}{\mu}\right) \qquad (5)$$

$$F(x) = \frac{\sqrt{Ex^2 + \frac{M}{\mu}x + \frac{1}{4}}}{E} - \frac{M/\mu}{2(-E)^{3/2}} \cos^{-1}\left(2Ex + \frac{M}{\mu}\right) \qquad (6)$$

For zero *binding* energy, $E=0$ ($M = \mu$):
$$\frac{\tau - \tau_0}{\mu} = G\left(\frac{R}{\mu}\right) - G\left(\frac{R_0}{\mu}\right) \qquad (7)$$

$$G(x) = \frac{2}{3}\left(x - \frac{1}{2}\right)\sqrt{x + \frac{1}{4}} \qquad (8)$$

For negative *binding* energy, $E>0$:
$$\frac{\tau - \tau_0}{\mu} = H\left(\frac{R}{\mu}\right) - H\left(\frac{R_0}{\mu}\right) \qquad (9)$$

$$H(x) = \frac{\sqrt{Ex^2 + \frac{M}{\mu}x + \frac{1}{4}}}{E} - \frac{M/\mu}{2(E)^{3/2}} \cosh^{-1}\left(2Ex + \frac{M}{\mu}\right). \qquad (10)$$

Note that the shell solution in terms of the dimensionless quantities $\tau/\mu$ and $R/\mu$ depends only on the dimensionless mass ratio, $M/\mu$.

Israel's parametric solutions can be obtained by the following parameterizations of $R$.

$$E<0:\ x = -\frac{M}{2E\mu} + \frac{1}{2E}\cos\Theta.$$

$$E=0:\ x = \lambda^2 - \frac{1}{4}.$$

$$E>0:\ x = -\frac{M}{2E\mu} + \frac{1}{2E}\cosh\Theta.$$

These solutions are plotted in Figure 1 for three sample values of the binding energy.

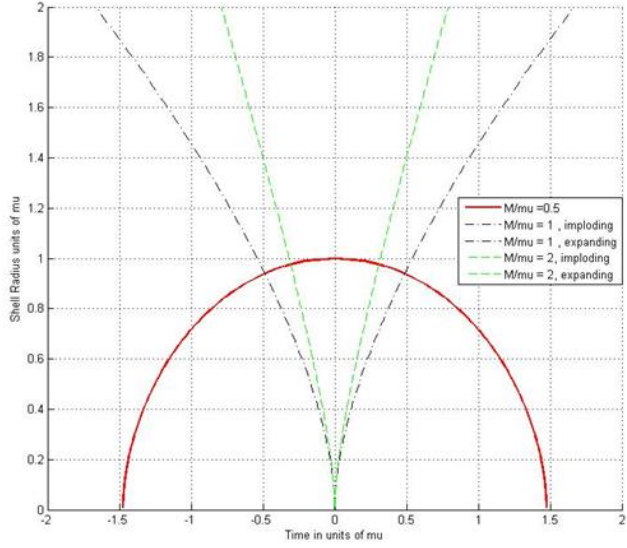

**FIG. 1 THIN SHELL SOLUTIONS FOR THREE ENERGY CASES**

## IV. Time dilation formulas

In this section we derive formulas for the time dilation of closely spaced clocks separated only by the gravitating thin shell.

From the embedding equations for the thin shell [1] we have:

$$\frac{dt}{d\tau} = \left[ f(R) + \dot{R}^2(\tau) \right]^{1/2} / f(R) \tag{11}$$

$$\frac{dT}{d\tau} = \left[ 1 + \dot{R}^2(\tau) \right]^{1/2} \tag{12}$$

Using equation (3) to eliminate $\dot{R}^2$ gives:

$$\frac{dt}{d\tau} = \frac{1}{f(R)} \left| \frac{M}{\mu} - \frac{\mu}{2R} \right| \tag{13}$$

$$\frac{dT}{d\tau} = \frac{M}{\mu} + \frac{\mu}{2R} \tag{14}$$

These two equations can be used to rewrite the metric in terms of the proper time on the shell:

$$ds^2 = -\left( \frac{M}{\mu} + \frac{\mu}{2R} \right)^2 d\tau^2 + dr^2 + r^2 d\Omega^2 \qquad \text{if } r < R(\tau) \tag{15}$$

$$ds^2 = -f(r) \left[ \frac{1}{f(R)} \left( \frac{M}{\mu} - \frac{\mu}{2R} \right) \right]^2 d\tau^2 + f(r)^{-1} dr^2 + r^2 d\Omega^2 \qquad \text{if } r > R(\tau) \tag{16}$$

We note that in this form the discontinuity in both $g_{\tau\tau}$ and $g_{rr}$ are readily apparent. This form of the metric uses the proper time of clocks on the shell for extension into the internal Minkowski region and into the external Schwarzschild region. The regions covered by equations (15) and (16) are the entirety of the manifold.

Returning to equation (14) we see that the relation between $T$ and $\tau$ is well behaved for all values of $R$. However, from equation (13), the relation between $t$ and $\tau$ still exhibits the coordinate singularity when $r<2M$ and $R<2M$. The standard analysis of the Schwarzschild spacetime geometry applies in this case.

From equations (15) and (16) we can derive the time dilation between inertial clocks at rest in the cavity and inertial clocks at infinity:

$$dT = \left(\frac{M}{\mu} + \frac{\mu}{2R}\right)d\tau$$

$$dt = \frac{1}{f(R)}\left|\frac{M}{\mu} - \frac{\mu}{2R}\right|d\tau$$

Taking the ratio yields:

$$dT = \frac{\left(\dfrac{M}{\mu} + \dfrac{\mu}{2R}\right)}{\left|\dfrac{M}{\mu} - \dfrac{\mu}{2R}\right|} f(R)dt$$

$$= \frac{\left(2MR + \mu^2\right)}{\left|2MR - \mu^2\right|} f(R)dt$$

This equation still depends upon the shell proper time due to the presence of $R$. The equation is only valid up to the time $\tau$ at which the shell crosses the event horizon, $R(\tau) = 2M$. In the following we consider only the cases for which $R(\tau) > 2M$.

For clocks that are at rest at $r_- = R(\tau) - \varepsilon$ inside the cavity and at $r_+ = R(\tau) + \varepsilon$ outside, where $\varepsilon$ is a small number, we have:

$$d\tau_- = dT = \left(\frac{M}{\mu} + \frac{\mu}{2R}\right)d\tau$$

$$d\tau_+ = \sqrt{f(r)}dt = \frac{\sqrt{f(r)}}{f(R)}\left|\frac{M}{\mu} - \frac{\mu}{2R}\right|d\tau$$

For this case we have the differential time dilation due to the thin shell:

$$d\tau_- = \frac{\left(2MR + \mu^2\right)f(R)}{\left|2MR - \mu^2\right|\sqrt{f(r_+)}}d\tau_+$$

In the limit $\varepsilon \to 0$ giving:

$$d\tau_- = \frac{\left(2MR + \mu^2\right)}{\left|2MR - \mu^2\right|}\sqrt{f(R)}d\tau_+ \qquad (17)$$

This result shows that there is a differential rate for clocks inside the cavity and those outside the cavity.

Equation (17) is plotted in Fig. 2 for several values of the mass ratio $a = M/\mu$. The cases displayed are for negative binding energy and apply to the imploding and exploding trajectories in Fig. 1. The plot only displays the case for $R(\tau) > 2M$.

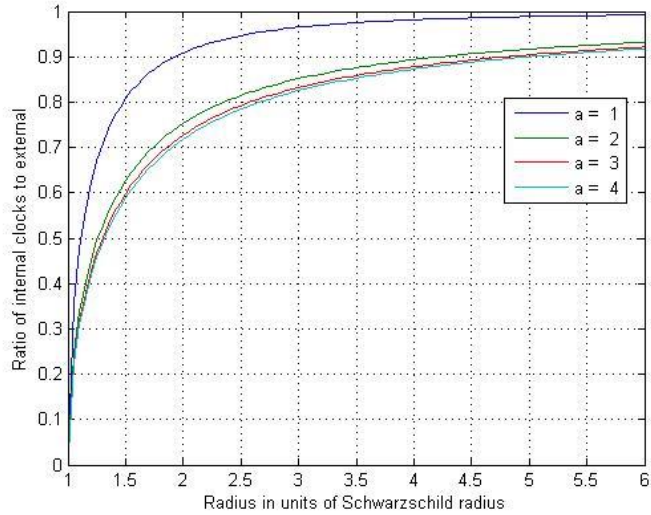

**FIG. 2 TIME DILATION OF CLOCKS SEPARATED BY A THIN SHELL**

## V. A geometric construction

The derivation of equation (17) proceeded in a straightforward algebraic approach. In this section we rederive the differential clock rates by geometric analysis based on the geometry inside and outside the shell.

In Fig. 3 we show two clocks positioned at constant coordinate distance from the center of the shell (as determined by the curvature of the shell). The diagram shows the shell's world line during the transition of the region between the clocks. Each edge is labeled by its invariant length.

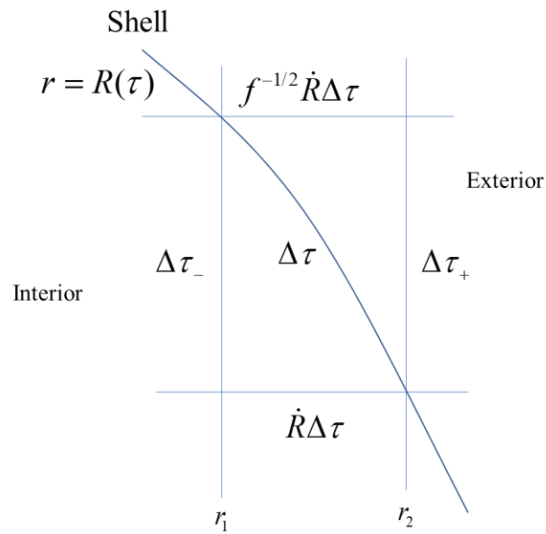

**FIG. 3 GEOMETRY OF THIN SHELL TIME DILATION**

From the diagram we obtain the following relations based on the common "hypotenuse" of the time-like segment of the shell with invariant proper time $\Delta \tau$:

$$-\Delta \tau^2 = -\Delta \tau_-^2 + \dot{R}^2 \Delta \tau^2$$
$$= -\Delta \tau_+^2 + f(\bar{r})^{-1} \dot{R}^2 \Delta \tau^2$$

The coordinate $\bar{r} = \frac{1}{2}(r_1 + r_2)$ and $\Delta \tau_-, \Delta \tau_+$ are the invariant lengths (i.e. proper time) of the clocks' world-lines. From the geometric intervals we obtain:

$$\Delta \tau_- = \sqrt{\Delta \tau^2 + \dot{R}^2 \Delta \tau^2} = \sqrt{1 + \dot{R}^2} \Delta \tau$$
$$\Delta \tau_+ = \sqrt{1 + f(\bar{r})^{-1} \dot{R}^2} \Delta \tau$$

Taking the ratio yields:

$$\Delta \tau_- = \frac{\sqrt{1 + \dot{R}^2}}{\sqrt{1 + f(\bar{r})^{-1} \dot{R}^2}} \Delta \tau_+$$

Upon using the EoM to eliminate $\dot{R}$ (cf. equation (3)) as in equations (13) and (14) we obtain equation (17) in the limit that $r_1 \to r_2$.

It is clear from this construction that $r_1$ and $r_2$ must lie outside the horizon, since the calculation assumes that curves at constant *r* values are time-like worldlines.

### VI.   Conclusions

In this paper we have derived time dilation formulae for observers in Israel's thin-shell solution. The derivation of time dilation between the regions of the space-time was derived both analytically and by geometric construction. The geometric construction provides a clear proof that there is a discontinuous time dilation between clocks at rest separated by the thin shell. We also derived the positivity constraint for the gravitational mass of the thin shell.

**Acknowledgement.** I wish to thank Werner Israel for reviewing an early version of this paper and his comments on the *M=0* case which has improved the paper.

[1] Israel, W., Singular hypersurfaces and thin shells in general relativity, *Il Nuovo Cimento* **44 B**(1), 1–14 (1966).

[2] Misner, Thorne and Wheeler (MTW), *Gravitation*, W.H. Freeman and Company, San Francisco, California, 1973.

[3] Bondi, H., Spherically symmetrical models in general relativity, *Monthly Notices of the Royal Astronomical Society* Nos. 5,6, p. 410 (1947).

[4] Synge, J.L., *Relativity: The general theory*, North-Holland Publishing Company, Amsterdam, 1971.

[5] Israel, W., Gravitational Collapse and Causality, *Phys. Rev.* **153**(5), 1388 (1967).